\documentstyle{mn}
\input{epsf}

\title{Towards gravitational-wave asteroseismology}

\author[Nils Andersson and Kostas D. Kokkotas]
{ Nils Andersson$^{1,2}$ and Kostas D. Kokkotas$^{3,4 }$\\ 
$^{1}$ Department of Physics,
Washington University, St Louis MO 63130, USA \\
$^{2}$ Institute f\"ur Theoretische Astrophysik, 
Universit\"at T\"ubingen,
D-72076 T\"ubingen, Germany\\
$^{3}$ Department of Physics, Aristotle University of Thessaloniki,
Thessaloniki 54006, Greece \\
$^{4}$ Max-Planck-Institut f\"ur Gravitationsphysik, Schlaatzweg 1,
14473 Potsdam, Germany}
\date{Accepted 1997  (?).
      Received 1997  (?);
      in original form  1997}
\begin{document}

\maketitle

\begin{abstract}
We present new results for pulsating neutron stars. We have calculated
the eigenfrequencies of the modes that one would expect to be the
most important gravitational-wave sources: the fundamental fluid $f$-mode,
the first pressure $p$-mode and the first gravitational-wave
$w$-mode, for twelve realistic equations of state. From this numerical
data we have inferred a set of ``empirical relations'' between the 
mode-frequencies and the parameters of the star (the radius $R$ and
the mass $M$). Some of these relation prove to be surprisingly
robust, and 
we show how they can be used to extract
the details of the star from observed modes. The results indicate
that, 
should the various pulsation modes be detected by the new generation
of 
gravitational-wave detectors that come online in a few years, the mass
and the radius of neutron stars can be deduced with errors 
no larger than a few percent.
\end{abstract}

\begin{keywords}
Stars : neutron - Radiation mechanisms: nonthermal
\end{keywords}

\section{Introduction}

\subsection{Motivation}

The day of the first undeniable detection of gravitational waves should not be 
far away. In less than five years at least five 
large interferometric 
gravitational-wave detectors (LIGO, VIRGO, GEO600 and
TAMA) will be operating. At the same time
a new generation
of spherical resonant detectors (GRAIL, SFERA etc) could be sensitive
enough to detect signals from supernova collapse
and binary coalescences in the Virgo cluster of galaxies.
In other words: Recent advancements in technology are heralding 
the era of gravitational-wave astronomy.
However, for this field to reach its full potential theoreticians 
must in advance point out the most promising sources, 
the optimal methods of detection and the appropriate bandwidth to
which the detectors should be tuned. 
Hence, the theoretical effort is presently focused on various sources
of potentially detectable gravitational waves, in order to 
both characterize the waves and device 
detailed detection strategies.
Once gravitational waves are detected the first 
task will be to identify the source. This should be possible
from the general character of the waveform and may not require
very accurate theoretical 
models, but accurate models will be of crucial importance 
for a deduction of the parameters of the
source. That is, for gravitational-wave ``astronomy''.

With this paper we contribute to this rapidly growing field in two ways. 
We present results for the gravitational waves from a pulsating 
relativistic star, 
eg. the  violent oscillations of a compact object formed
after a core collapse. These results provide a means for taking the
fingerprints of the source, and suggest optimal bandwidths 
to which a detector should be tuned to enable  
detection of such signals.
Specifically, we discuss how the information carried by 
gravitational waves from a pulsating
star can be used to infer, with good precision,
both the mass and the radius of the star: Data that would strongly
constrain the supranuclear equation of state (EOS). 

The idea behind the present work is a familiar one in astronomy:
For many years, studies of the light
variation  of variable stars have been used to deduce their internal
structure \cite{unno}. 
The Newtonian theory of stellar pulsation was
to a large extent 
developed in order to explain the pulsations of Cepheids and RR Lyrae.
This approach, known as Asteroseismology (Helioseismology
in the specific case of the Sun), has been quite successful in
recent years.
The relativistic theory of 
stellar pulsation has now been developed for thirty years, but it has not yet
been applied in a similar way. 
So far, the relativistic theory has no immediate 
connections to observations 
(that are not already
provided by the Newtonian theory). We believe that this situation
will change once the gravitational-wave
window to the universe is opened, and with this article we discuss how
the information carried by the gravitational-wave signal can be
inverted to estimate the parameters of pulsating stars. That is,
we take the first (small) step towards gravitational-wave
asteroseismology.

\subsection{Detectability of the waves}

At the present time it is not clear that the gravitational waves
from pulsating neutron stars will be seen by the detectors
that are presently under construction. Our relative ignorance  
in this matter is due to the lack of 
accurate, fully relativistic, models 
of, for example, the gravitational collapse that follows
a supernova. At present we simply do not know how much energy 
will be radiated through the oscillation modes
of a nascent neutron star.
But one can argue that
the released energy could be considerable. One generally
expects the newly formed neutron star to pulsate 
wildly during the first few seconds following the collapse.
This pulsation will be damped mainly through gravitational waves.
This means that the signal,
that carries the signature of the collapsed
object, may be invisible in the electromagnetic spectrum, but 
the amplitude of the emerging gravitational waves could be
considerable. 
The energy stored in the pulsation can potentially
be of the same order as the kinetic energy of the collapse.
In fact, it is not unreasonable to expect that
a significant part of the mass energy
of the newly formed object is radiated in this way.

The spectrum of a pulsating relativistic star is tremendously rich, 
since essentially every feature of the star can be directly associated
with (at least) one distinct family of pulsation modes \cite{mcd,akk96}. 
But it seems likely that  
only a few of these modes will carry away the bulk of 
the radiated energy \cite{aaks}.
From the gravitational-wave point of view, the most important modes
are the fundamental ($f$) mode of fluid oscillation, 
the first and maybe the second pressure ($p$) modes 
 and the first gravitational-wave ($w$) mode
\cite{ks92,aks,akk96}.
Other pulsation modes, eg. the $g$ (gravity),
higher order $p$-modes,   
$s$ (shear), $t$ (toroidal), $i$ (interface) modes, can be accounted for    
with  Newtonian dynamics since they do not emit  
significant amounts of gravitational radiation \cite{mcd}. 
For a historical description of, and further details on, 
the theory of relativistic 
stellar pulsation we refer the reader to a recent review article by
one of us \cite{kokkotas97}. A detailed discussion of the relativistic
perturbation equations was recently provided by Allen et al.
\shortcite{aaks}.

The pulsation modes of a neutron star are likely to be
excited in many dynamical processes, but will the resultant
gravitational waves be strong enough to be detectable on Earth?
As already mentioned, the answer to that question is unclear and
demands more accurate modeling, but
it is straightforward to derive
useful order of magnitude estimates. As we have have recently shown
elsewhere \cite{ak96},  
we get  
\begin{equation}
h_{\rm eff} \sim 2.2\times 10^{-21} \left( {E\over 10^{-6} M_\odot c^2}
 \right)^{1/2} \left(  { 2 {\rm kHz} \over f } \right)^{1/2}
\left( {50{\rm kpc} \over r} \right) \ ,
\end{equation}
for the $f$-mode, and
\begin{equation}
h_{\rm eff}\sim 9.7\times 10^{-22} \left( { E \over 10^{-6} M_\odot c^2}
 \right)^{1/2} \left( { 10 {\rm kHz} \over f } \right)^{1/2}
\left( { 50{\rm kpc} \over r } \right) \ ,
\end{equation}
for the fundamental $w$-mode.
Here we  have used typical parameters for the pulsation modes, 
$E$ is the available pulsation energy, 
and the distance scale used is that to SN1987A.
In this volume of space one would not expect to see more than 
one event per ten years or so.  
However, the assumption that the energy released through 
gravitational waves in a supernova is of the order of 
 $10^{-6} M_\odot c^2$ is probably conservative. 
If 
a substantial fraction of the binding energy of a neutron star 
were released through the pulsation modes they could 
potentially be observable all the way out to the Virgo cluster.
Then we could hope to see several events per year.

Suppose we want to know how much energy must go into
each mode to achieve an effective gravitational-wave amplitude
$h_{\rm eff}\sim 10^{-21}$ (the order of magnitude  a
signal must have to be ``detectable'') 
when the source is at a distance of 15 Mpc.
We can invert the above relations, and find that at least 1\% of a 
solar mass must be radiated through these modes if they are to be 
detectable at the Virgo distance. The specific numbers are
$E\approx 0.019 M_\odot c^2$ for the $f$-mode and $E\approx 0.096 M_\odot
c^2$ for the $w$-mode. To assume that this amount of energy
actually goes into these modes seems somewhat optimistic, but
the possibility should not be ruled out. Anyway, the rough estimates
indicate that the pulsations of a nascent neutron
star in the local group of galaxies could well be detectable.
This may not be a very frequent event, but as we shall see the pay-off
of its detection could be great.

\subsection{Addressing the inverse problem}

Considering the possibility of a future detection 
it is relevant to pose the 
``inverse problem'' for gravitational waves from pulsating stars.
Once we have observed the waves, can we deduce the details of the star
from which they originated?
To answer this question we have    
calculated the frequencies
and damping times of the modes
that we
expect to lead to the strongest gravitational waves for a selection of
EOS. 
The numerical method used for the calculation 
is essentially that described by Andersson, Kokkotas and
Schutz \shortcite{aks}. 
The study includes several 
stellar models for each EOS. The obtained data, which is tabulated 
in Appendix A, extends
previous results of Lindblom and Detweiler \shortcite{ld} in two ways:
i) a few  modern EOS are included, and ii) we have added
results both for  $p$- and $w$-modes.
We have used this numerical data to
create useful ``empirical'' relations between the
``observables'' (frequencies and damping times) and the parameters
of the star (mass, radius and possibly the EOS).
As we will show in the following, these relations can be used to infer the 
stellar parameters from detected mode data.
 
In this study we  have not taken into 
account the effects of rotation. There are two reasons for this:
Firstly,  rotation should
have a marginal effect, except for in the most rapidly 
spinning cases, since rotational effects scale as the angular velocity squared.
Secondly, and more importantly,
there is at present no available method that can be used to
study  pulsation modes of a rapidly rotating relativistic star.
Such methods must be developed, and once the relevant results become available
the present study can be complemented to incorporate them.

Before we proceed to discuss the deduced empirical relations, a
few comments on our choice of stellar models are in order.
In Appendix A we tabulate data for various oscillation modes 
($f$, $p$ and $w$) for  twelve different EOS listed in Table~\ref{table1}.
The chosen pulsation modes are those that i) should produce the strongest
gravitational waves, and ii)
lie inside the 
bandwidth of the detectors which are either planned or under 
construction. Most of the 
 EOS were taken from the old Arnett and Bowers catalogue 
\cite{ab74}. Although some of these EOS might be outdated
none of them are ruled out by present
observations.
Furthermore, the range of stiffness of the EOS listed by Arnett and
Bowers is still relevant today.  This is important for the present
study: In order for our analysis to be robust it is necessary that our
sample of EOS spans the anticipated range of stiffness.  
However, we have also included three more
modern EOS: one of the models of Wiringa et al
\shortcite{WFF}  and
two models from Glendenning \shortcite{NKG}.  
For the EOS that were also considered by  
Lindblom and Detweiler \shortcite{ld} we have chosen identical
stellar models to facilitate a comparison of the results. 
Finally, 
we have only included stellar models whose masses and 
radii are within the limits accepted by current observations 
\cite{finn,Kerk95}. 

\begin{table}
  \caption{The twelve Equations of State that were included in the study.}
  \begin{tabular}{@{}ll}
 Model  & Reference \\ 
 A      & Pandharipande \shortcite{Pand71} (neutron) \\
 B      & Pandharipande \shortcite{Pand71} (hyperonic; model C)\\
 C      & Bethe and Johnson \shortcite{BJ74} (model I) \\ 
 D	& Bethe and Johnson \shortcite{BJ74} (model V)\\
 E	& Moszkowski \shortcite{Moszk74}\\
 F	& Arponen \shortcite{Arpon72}\\
 G	& Canuto and Chitre \shortcite{CC74}\\
 I	& Cohen et al. \shortcite{Cohen70}\\
 L	& Pandharipande et al. \shortcite{Pandh76}\\
 WFF	& Wiringa et al.\shortcite{WFF}\\
 G$_{240}$	& Glendenning \shortcite{NKG} K240\\
 G$_{300}$	& Glendenning \shortcite{NKG} K300 \\
\end{tabular}
\label{table1}\end{table}

\section{What can we learn from observations?}

Our present understanding of neutron stars comes mainly from 
X-ray and radio-timing observations. 
These observations provide some insight into the 
structure of these objects and the properties of supranuclear
matter. The most commonly and accurately observed parameter
is the  rotation period, and we know that radio pulsars can spin 
very fast (the shortest observed period being the 1.56 ms of PSR
1937+21). 
Another   
basic observable, that can be obtained (in a few cases)
with some accuracy from todays observations,
is the mass of the neutron star. 
As Finn \shortcite{finn} has shown, the observations of radio pulsars
indicate that $1.01 < {M/M_\odot} < 1.64$. Similarly, van der Kerkwijk
et al \shortcite{Kerk95} find that data for X-ray pulsars  indicate
$1.04 < {M/M_\odot} < 1.88$. The data used in these two studies 
is actually consistent with (if one includes error bars)
$M<1.44 M_\odot$. 
We now recall that the various
EOS that have been proposed by theoretical physicists 
can be divided into two major categories: i) the ``soft'' EOS which
typically lead to neutron star models with maximum masses around 
$1.4 M_\odot$ and radii usually smaller than 10 km, and ii)
  the ``stiff'' EOS
with the maximum values  $M\sim 1.8 M_\odot$ and $R\sim 15$ km 
\cite{ab77}.
From this one can deduce that, even though the constraint put on the
neutron star mass by present observations seems strong, 
it does not rule out many
of the proposed EOS.
In order to arrive at a more useful result we are
likely to need detailed observations also of the stellar radius. 
Unfortunately, available data provide
little information about the radius. The recent observations of
quasiperiodic oscillations in low mass X-ray binaries indicate
that $R<6M$, but again this is not a severe constraint. 
Although a number of attempts have been made, using either X-ray observations 
\cite{lewin} or the limiting spin period of neutron stars \cite{fip}, 
to
put constraints on the mass-radius relation,
we do not yet have a method
which can provide the desired answer. 

\subsection{A detection scenario}

In view of this situation, any method that can be used to infer
neutron star parameters is a welcome addition. 
Of specific interest may be the new possibilities
offered once gravitational-wave observations become reality. 
An obvious question is to what extent one can solve the inverse
problem in gravitational-wave astronomy. In this paper we address
this issue for the case of waves from pulsating neutron stars. 
In Appendix A we provide extended tables with frequencies and
damping times for the most relevant pulsation modes (as far as
gravitational waves are concerned) of various stellar models 
created from a range of realistic EOS.
We will suppose that these modes can be detected by some
future generation of gravitational-wave detectors, and
investigate to what precision we can hope to calculate the 
parameters of the source from observed data.
 
Let us suppose that a nearby supernova explodes, say 
in the Local Group of galaxies, and is
followed by a core collapse that leads to the formation of 
a compact object. As the dust from the collapse settles the 
compact object pulsates wildly in its various oscillation modes,
generating a gravitational-wave signal which  
is composed of an overlapping of different frequencies.. 
We will assume that the results of Allen et al \shortcite{aaks}
can be brought to bear on this situation, i.e. that  most of the energy 
is radiated through the $f$-mode, a few $p$-modes and the first
$w$-mode. Our detector picks up this signal, and a subsequent 
Fourier analysis of the data stream yields the frequencies 
and the energy content in each mode.
 
The first question to be answered by the gravitational-wave
astronomer concerns what kind of compact object could produce the
detected signal. Is it a
black hole or a neutron star? The pulsation of  these  objects
lead to qualitatively similar gravitational waves, eg. exponentially
damped oscillations, but the      
question should nevertheless be relatively easy to answer.
If more than one of the stellar pulsation modes is observed the answer
is clear, but even if we only observe only one single mode the two
cases should be easy to distinguish.
The fundamental (quadrupole) 
quasinormal mode frequency of a Schwarzschild black
hole follows from 
\begin{equation}
f \approx 12 {\rm kHz } \left( {M_\odot \over M} \right) , 
\end{equation}
while the associated e-folding time is
\begin{equation}
\tau \approx 0.05 {\rm ms } \left( {M \over M_\odot} \right) \ .
\end{equation} 
That is, the oscillations of a 10 $M_\odot$ black hole lie in 
the frequency range of the $f$-mode for a typical neutron star (see
Appendix A). But the two signals will differ greatly in 
 the damping time, the e-folding time of 
the black hole being nearly three orders of magnitude shorter than 
that of the neutron star $f$-mode.

Having excluded the possibility that our signal came from a
black hole, we want to know  the mass and the radius of 
the newly born neutron star. We also want to decide which 
of the proposed EOS that best represents 
this star. To address these questions we can use a set of empirical
relations deduced from the 
data of  Appendix A: Relations that can be used to 
estimate the mass,
the radius and the EOS of the neutron star with good precision.

\subsection{Empirical relations}

\begin{figure}
\centerline{\epsfxsize=9cm \epsfysize=10cm \epsfbox{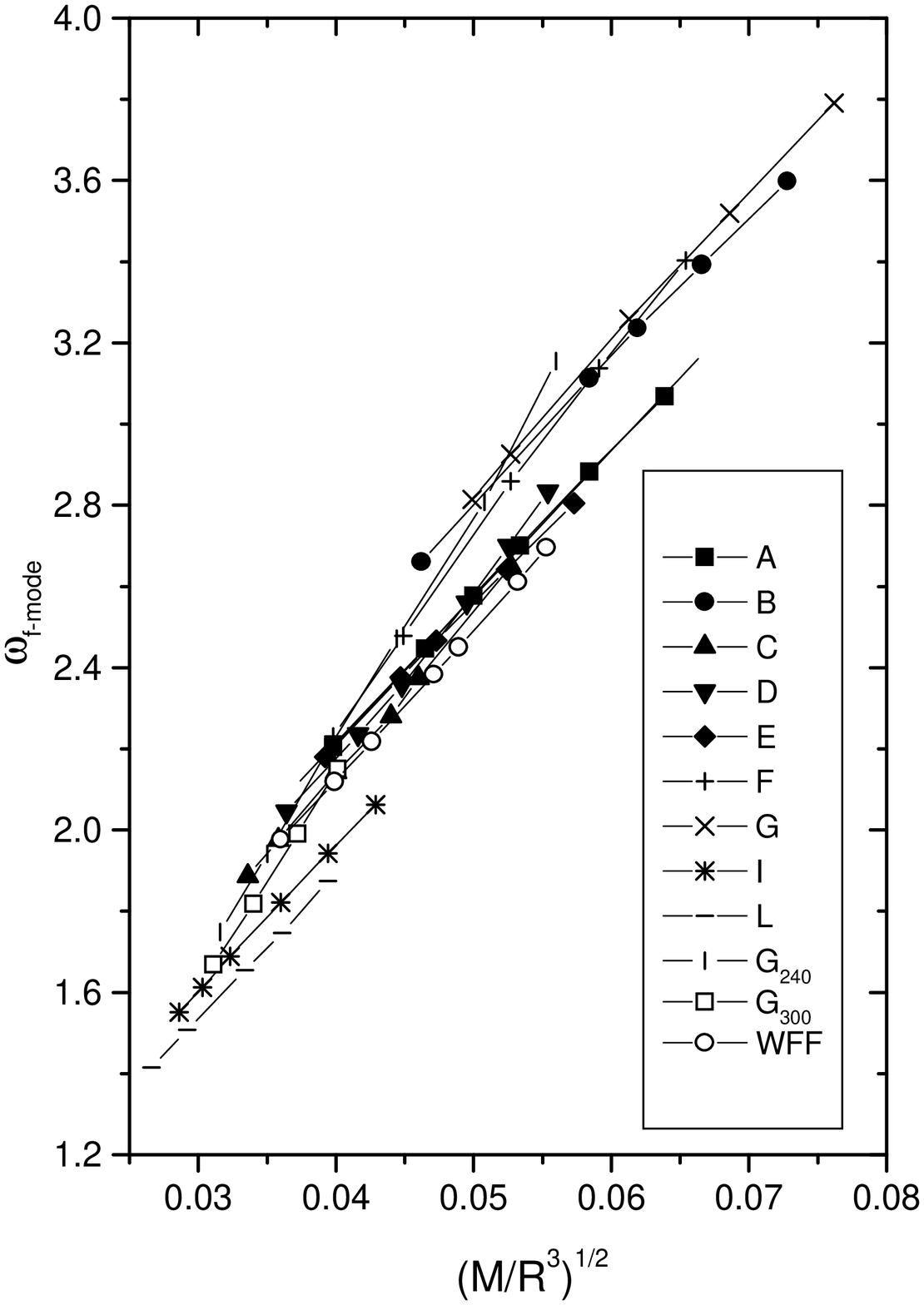}}
\caption{
The numerically obtained
$f$-mode frequencies plotted as functions of the mean stellar density 
($M$ and $R$ are in km and $\omega_{f-mode}$ in kHz).
}
\label{ffw}\end{figure}

Let us first consider the frequency of the $f$-mode.
It is well known that the characteristic time-scale of any dynamical
process
is related to the mean density of the mass involved 
(see Misner, Thorne and Wheeler \shortcite{MTW} ch. 36.2). 
This notion should be relevant for the 
fluid oscillation modes of a star, and  we consequently expect that
$\omega_f \sim \bar{\rho}^{1/2}$. That is,    
we should normalize the 
$f$-mode frequency with the average density of the star. 
The result of doing this is shown 
in Figure~\ref{ffw}. 
From this Figure it is apparent that the relation between 
the $f$-mode frequencies and the mean density is almost linear, 
and a  linear fitting leads to the following simple relation: 
\begin{equation}
\omega_f (kHz) 
\approx 0.78 + 1.635 \left(  {{\bar M} \over {{\bar R}^3}} \right)^{1/2}
\ ,
\label{rfw}
\end{equation}
where we have introduced the dimensionless variables
\begin{equation}
{\bar M} = {M \over {1.4 M_\odot}} \quad \mbox{and} \quad
{\bar R} = {R \over 10 {\rm km} } \ . 
\end{equation}
>From equation (\ref{rfw}) follows that the typical  $f$-mode frequency
is around 2.4 kHz.

\begin{figure}
\centerline{\epsfxsize=9cm \epsfysize=10cm \epsfbox{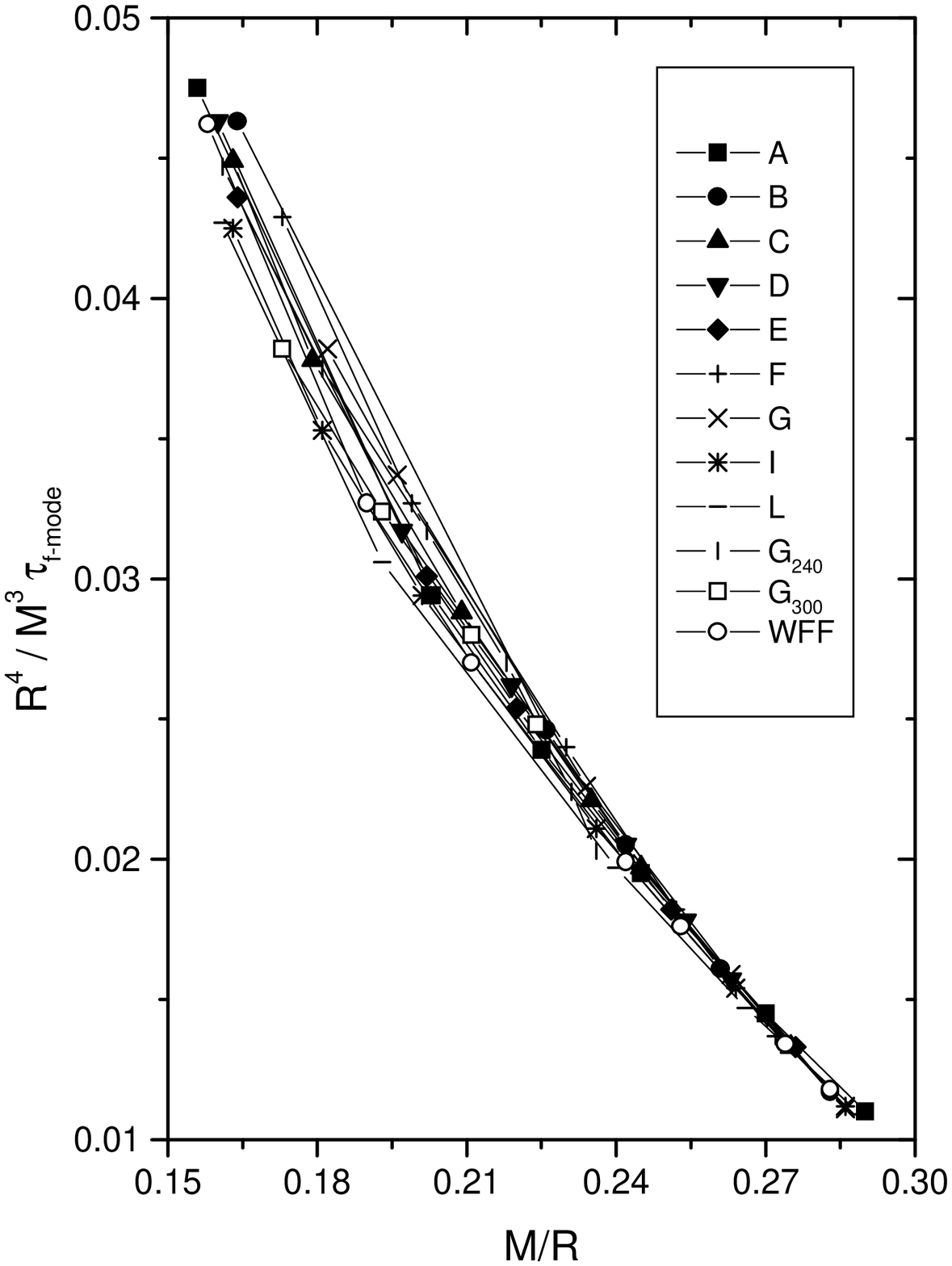}}
\caption{ The normalized damping time of the $f$-modes as function
of the stellar compactness 
($M$ and $R$ are in km and $\tau_{f-mode}$ in secs). }
\label{fftau}\end{figure}

To deduce a corresponding relation for the damping rate 
of the $f$-mode, we can use the rough estimate given by the
 quadrupole formula. That is, the damping time should follow from
\begin{equation}
\tau_f \sim {\mbox{oscillation energy} \over{ \mbox{power emitted in GWs}}}
\sim R\left({R\over M}\right)^3 \ .
\end{equation}
Using this normalization we plot the functional
$(\tau_f M^3/R^4)^{-1}$ as a function of the stellar compactness, cf. 
Figure~\ref{fftau}. 
The data shown in this Figure leads to a
relation between the damping time of the $f$-mode and the stellar parameters
$M$ and $R$:
\begin{equation}
{1 \over \tau_f ({\rm s})} \approx {{\bar M}^3 \over {\bar R}^4} 
\left[ 22.85 - 14.65 \left( {{\bar M}\over {\bar R}} \right)\right] \ .
\label{rftau}
\end{equation}
The small deviation of the numerical data
from the above formula is apparent in
Figure~\ref{fftau}, and one can easily see that a typical value for 
the damping time of the $f$-mode is a tenth of a second.

\begin{figure}
\centerline{\epsfxsize=9cm \epsfysize=10cm \epsfbox{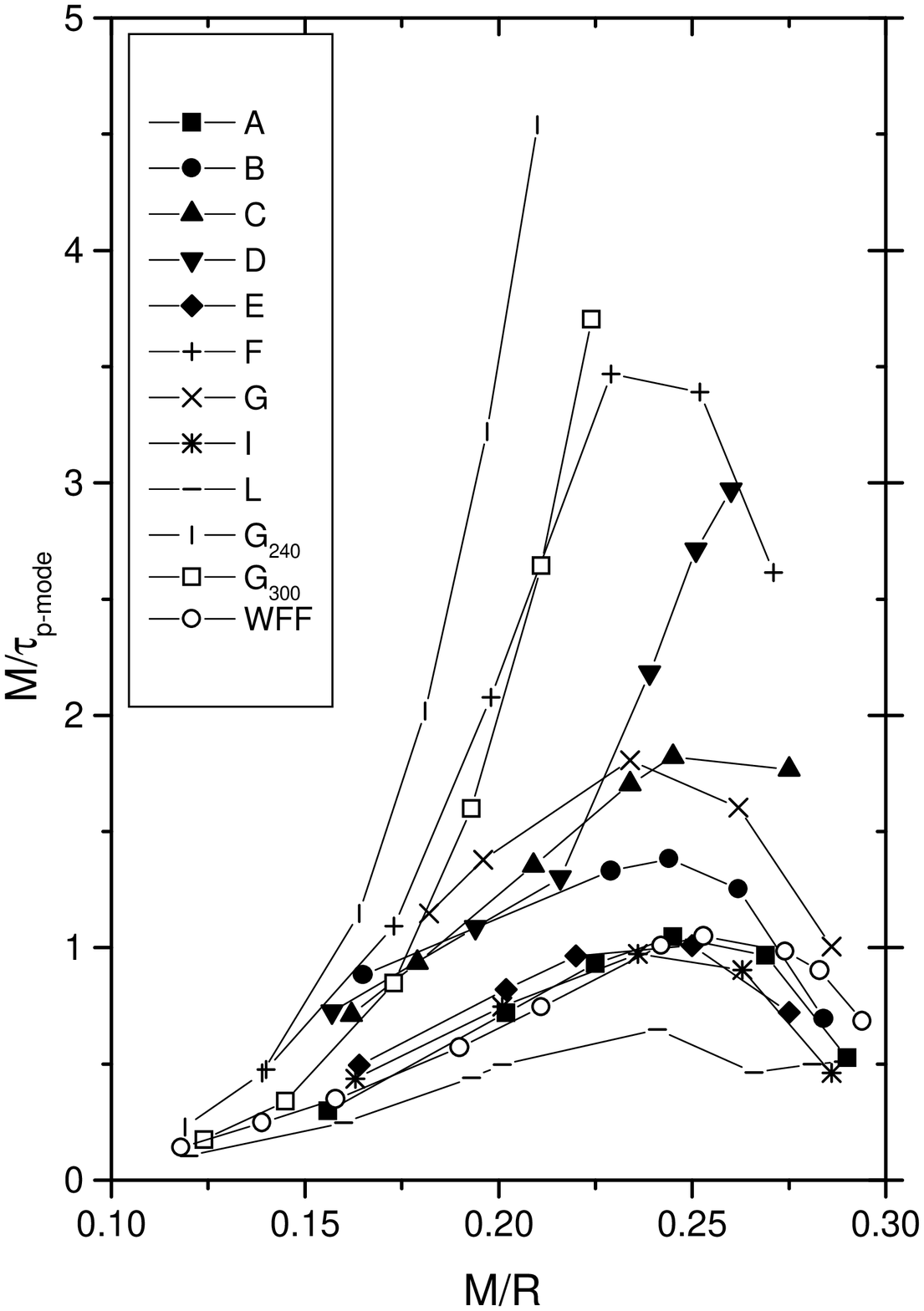}}
\caption{ The damping time of the $p$-modes as function
of the stellar compactness 
($M$ and $R$ are in km and $\tau_{p-mode}$ in secs). }
\label{fptau}\end{figure}

For the damping rate of the $p$-modes the situation is not so 
favorable. This is because the damping  of the $p$-modes 
is more sensitive to changes in the modal distribution inside the
star. Thus, different EOS lead to rather different $p$-mode 
damping rates, cf. Figure~\ref{fptau}.  
Previous evidence for polytropes \cite{akpoly} actually indicate that this
would be the case. Clearly, an empirical relation based on the data
in Figure~\ref{fptau} would not be very robust.

\begin{figure}
\centerline{\epsfxsize=9cm \epsfysize=10cm \epsfbox{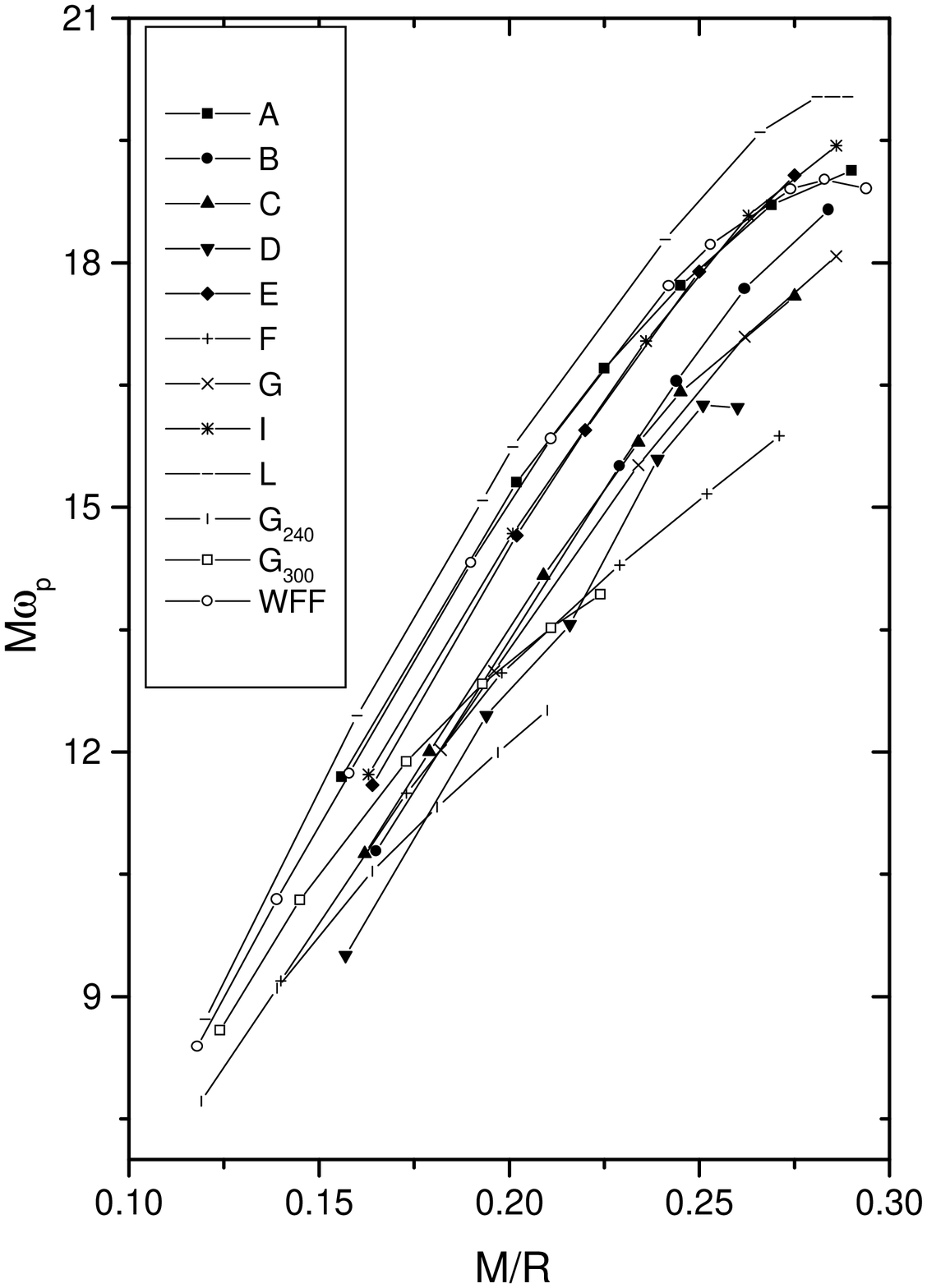}}
\caption{
The $p$-mode frequencies plotted a function of the compactness of the star
($M$ and $R$ are in km and $\omega_{p-mode}$ in kHz).}
\label{fpw}
\end{figure}

The situation is slightly better if we consider the oscillation
frequency
of the $p$-mode. From the data graphed in  
Figure~\ref{fpw}
we can deduce a relation
between the $p$-mode frequency and the parameters of the star: 
\begin{equation}
\omega_p ({\rm kHz}) 
\approx {1\over {{\bar M}}}\left(1.75 + 5.59 {{\bar M} \over {{\bar R}}} \right)
\ ,\label{rpw}
\end{equation}
and we see that a typical
 $p$-mode frequency is around 7 kHz. Although 
the data for several EOS deviate significantly from (\ref{rpw})
it is still a useful result. Stellar masses and radii deduced from it
will not be as accurate as ones based on $f$-mode
data, but on the other hand, if  
$M$ and $R$ are obtained in some other way (say, from a combination
of observed $f$- and $w$-modes) the $p$-mode can be used to
deduce the relevant EOS.  
  
\begin{figure}
\centerline{\epsfxsize=9cm \epsfysize=10cm \epsfbox{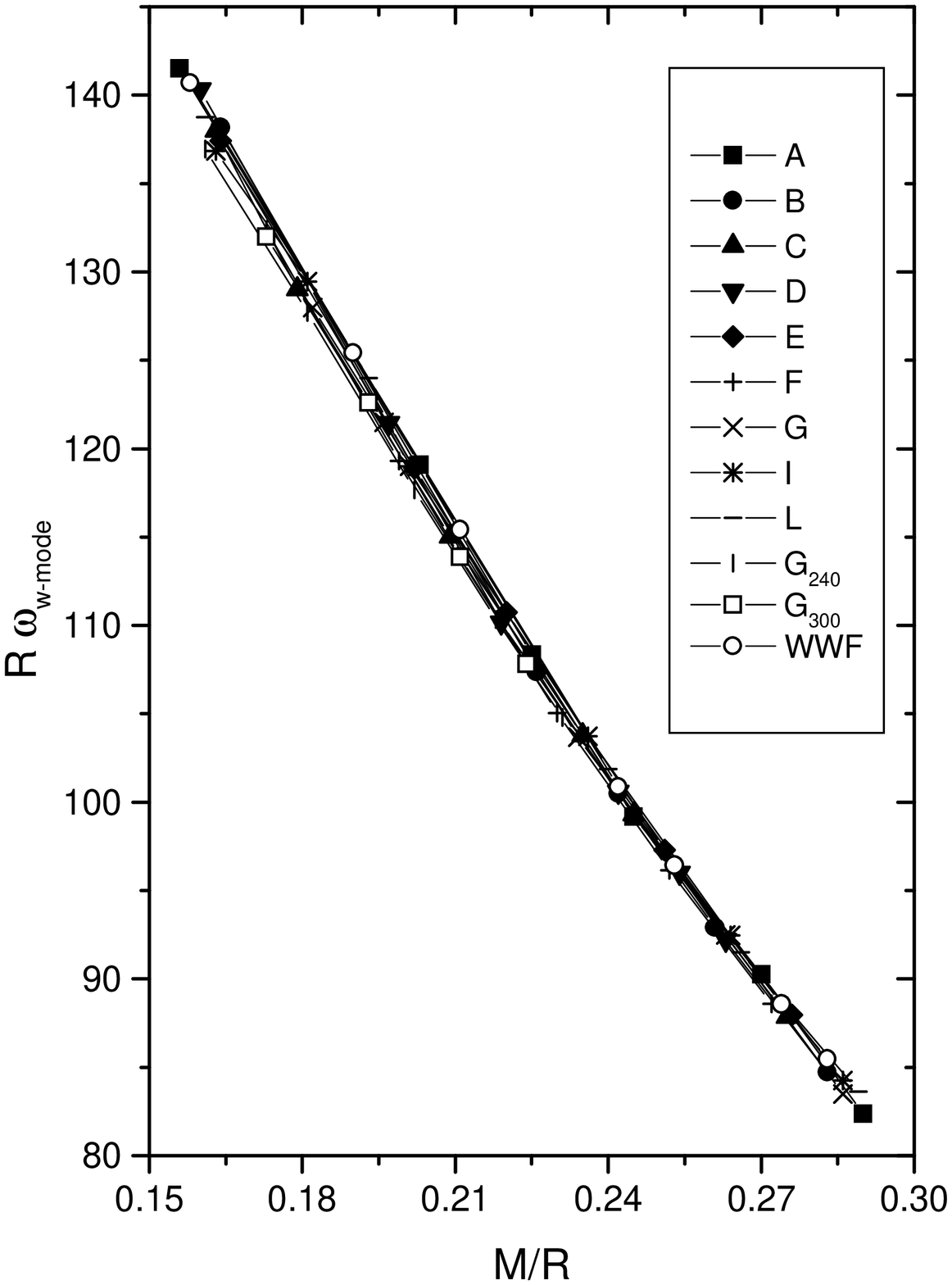}}
\caption{
The functional $R \omega_w$ as function of the compactness of the star
($M$ and $R$ are in km and $\omega_{w-mode}$ in kHz).
}
\label{fww}
\end{figure}

\begin{figure}
\centerline{\epsfxsize=9cm \epsfysize=10cm \epsfbox{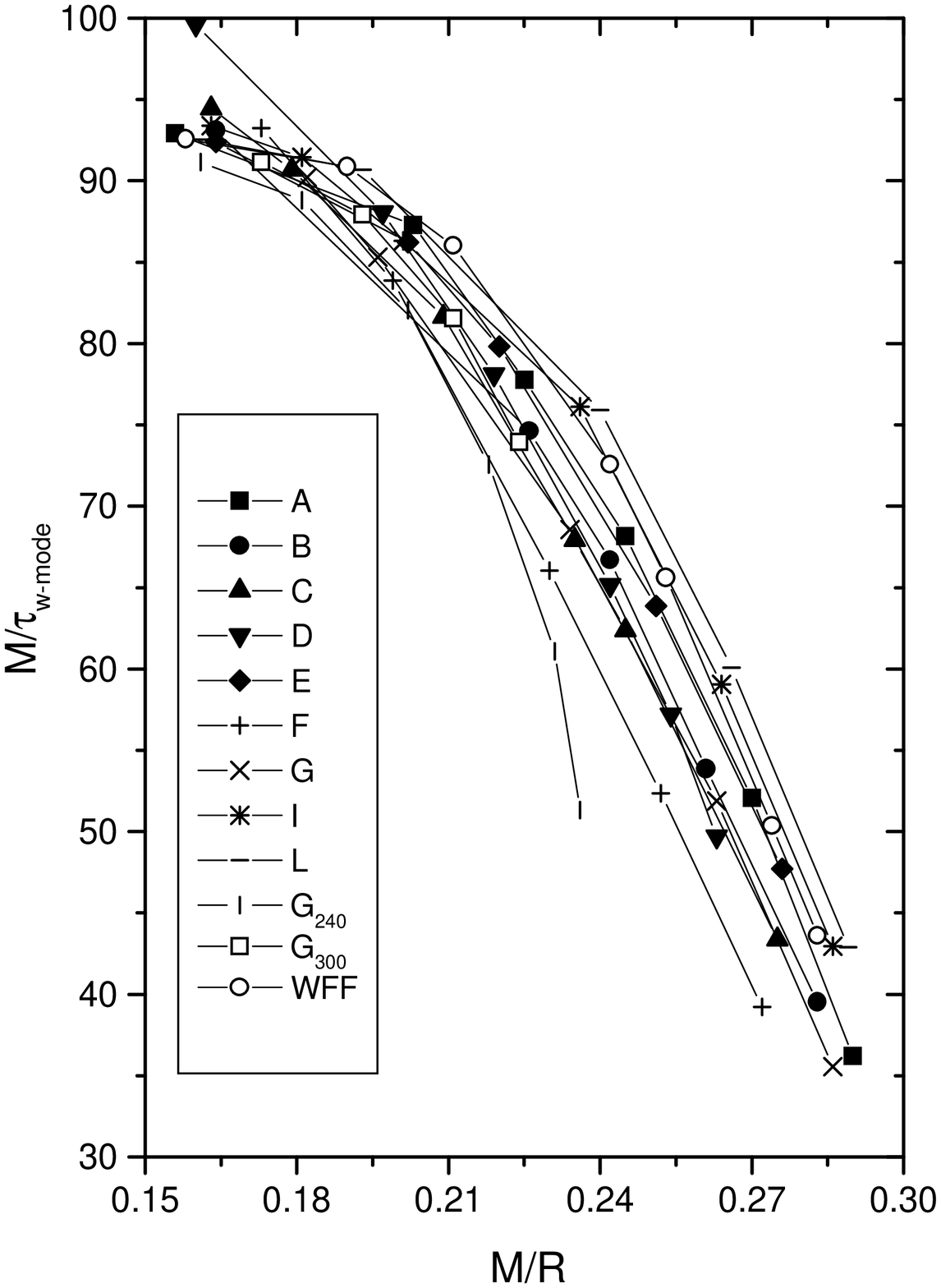}}
\caption{
The functional $M /\tau_w$ as  function of the compactness of the star
($M$ and $R$ are in km and $\tau_{w-mode}$ in msecs).
}
\label{fwtau}
\end{figure}

That empirical relations based on $p$-mode data would be less
robust and useful than those for the $f$-mode was expected, since 
the $p$-modes are sensitive to changes in the 
matter distribution inside the star. In contrast, the
gravitational-wave $w$-modes  should lead to very robust results.
It is well known 
\cite{ks92,akk96} that the $w$-modes do not excite a significant 
fluid motion. Thus, they are 
more or less independent of the characteristics of the fluid:
The frequencies do not depend on the sound speed and the damping
times cannot be modeled by the quadrupole formula.
But we can nevertheless deduce the appropriate
normalization 
for the $w$-mode data listed in Appendix A. 
Analytic results for model problems
for the $w$-modes \cite{ks86,nils96}, 
show that the frequency
of the $w$-mode is inversely proportional to the size of the star.
This is clear from the data in Figure~\ref{fww}.  
Meanwhile, the 
damping time is related to
the compactness of the star, i.e. the more relativistic the star is the longer
the $w$-mode oscillation lasts. This is shown in Figure~\ref{fwtau}.
These properties have already been discussed in some detail by
Andersson, Kojima and Kokkotas \shortcite{akk96}
for uniform density stars.

For the present numerical data (shown in Figures~\ref{fww} and
\ref{fwtau}) 
we find 
the following relations for the frequency and damping of the first $w$-mode:
\begin{equation}
\omega_w ({\rm kHz}) \approx {1\over {\bar R}}
         \left[ 20.92 - 9.14 \left( {\bar M} \over {\bar R} \right)
\right] \ ,
\label{rww}
\end{equation}
and
\begin{equation}
{1 \over \tau_w ({\rm ms})} \approx {1 \over {\bar M}} 
\left[ 5.74 + 103 \left( {{\bar M}\over {\bar R}} \right)
- 67.45 \left( {{\bar M}\over {\bar R}} \right)^2 \right] \ .
\label{rwtau}
\end{equation}
We see that a typical value for the $w$-mode frequency is 12 kHz, but
since the frequency depends strongly on the radius of the star
it varies greatly for different EOS. 
For example, for a very stiff EOS (L) the 
$w$-mode frequency is around 6 kHz while for the
softest  EOS in our set (G) the typical frequency is around 14 kHz.
The $w$-mode damping time is comparable to that of an oscillating 
black hole with the same mass,  i.e. it is typically less than a tenth 
of a millisecond.

\subsection{A simple experiment}

In principle, 
the relations we have deduced between the various pulsation modes and the
stellar parameters can be used to infer $M$ and $R$ (or some
combination thereof) from detected mode-data.    
The five relations (\ref{rfw}) --- (\ref{rwtau}) 
form an over-determined system of five equations
for the two unknown quantities $R$ and $M$. One would expect
this system to
 provide an accurate characterization of the star
in the ideal case when the gravitational-wave 
signal carries energy in all modes 
($f$, $p$ and $w$). 

This idea is promising and simple enough, but we need to examine
how well it can work in 
practice. To do this we have constructed a set of independent 
polytropic stellar models ($p = K \rho^{1+1/N}$) with varying polytropic 
indices ($N=0.8;1;1.2$). We have determined the $f$-mode, the first
$p$-mode and
the slowest 
damped  $w$-mode for each of these models. 
We let this data
represent ``observed''
gravitational-wave signals, and use 
 various combinations of the
relations (\ref{rfw}) -- (\ref{rwtau}) to extract the values of the 
masses and radii of the stellar models. 

\begin{figure}
\centerline{\epsfxsize=9cm \epsfysize=10cm \epsfbox{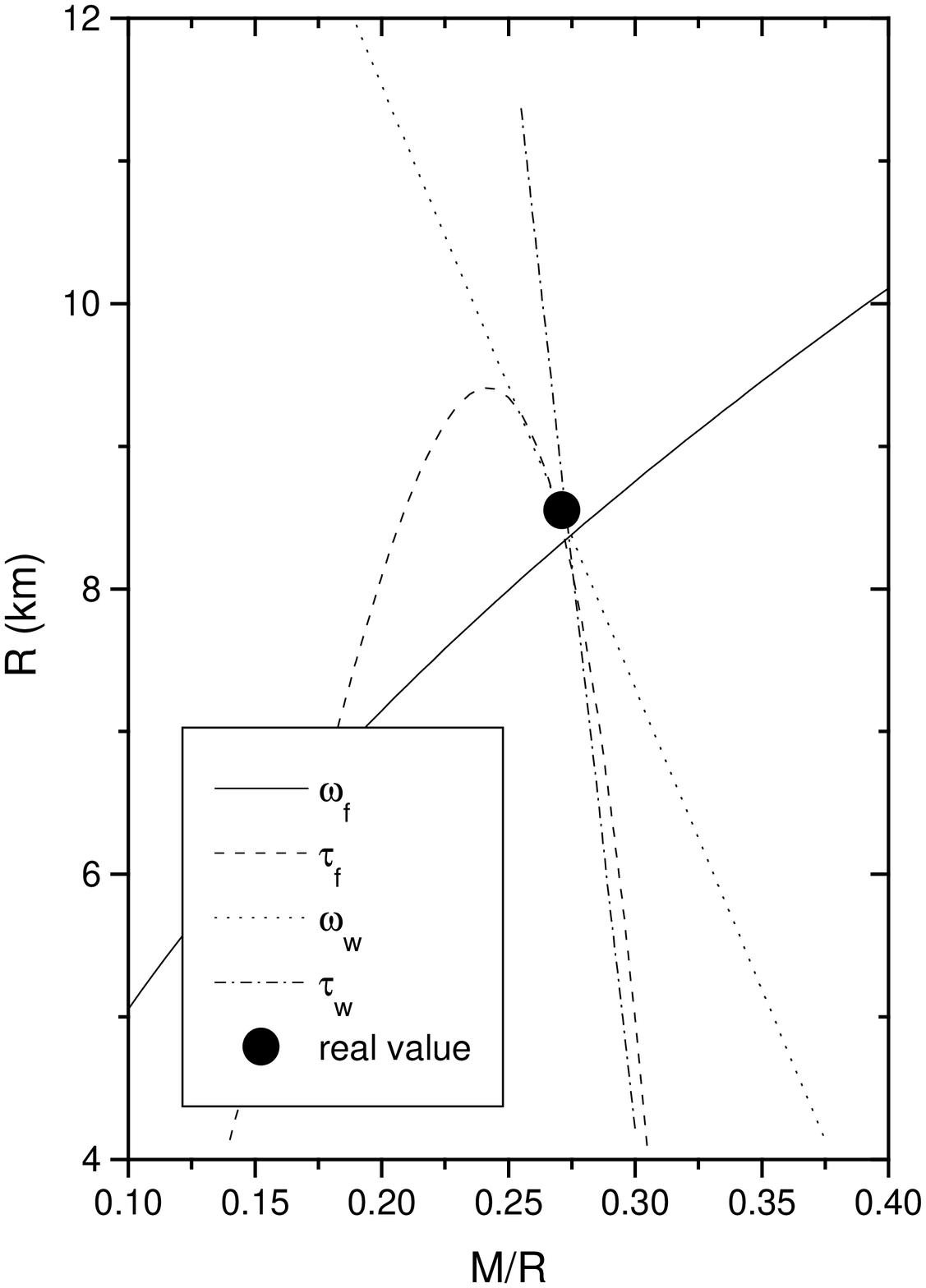}}
\caption{
An illustration of how accurately the radius and the mass of a
star can be inferred from detected mode data and our empirical relations.
}
\label{fig5}
\end{figure}

In Figure~\ref{fig5} we show the
result of a combination of the relations for $f$- and $w-$modes 
[(\ref{rfw}), (\ref{rftau}), 
(\ref{rww}) and (\ref{rwtau})] for one of the polytropic models.
In the Figure, a filled
circle represents the true parameters of the star, and it is clear that
estimates based on the above relations can be very accurate.
More detailed results are listed in 
Table 2. The typical errors of a parameter estimation based on the 
oscillation frequencies of the $f$- and the $w$-mode  
[the combination of (\ref{rfw}) and (\ref{rww})] 
are  (5\% ,2\%) where 
the first number is the error in the radius and the second 
is the error in the mass. 
Combining the frequency and damping of the $f$-mode [(\ref{rfw}) and
 (\ref{rftau})] we find  (6.5\% ,17.6\%). The
$f$-mode frequency and the $w$-mode damping rate 
[(\ref{rfw}) and  (\ref{rwtau})] lead to (5.6\% ,1.4\%),
while the $w$-mode frequency and the $f$-mode damping  
[(\ref{rww}) and (\ref{rftau})] yield  (3.2\% ,1.9\%).
A combination of the $w$-mode frequency and damping rate   
[(\ref{rww}) - (\ref{rwtau})] leads to  (3.9\% ,1.6\%).
Finally, by combining the damping rates of the $f$- and the $w$-mode   
[(\ref{rwtau}) and (\ref{rftau})] we get  (2.1\% ,6.3\%).
These results are rather impressive. The robustness of our empirical 
relations for $f$- and $w$-modes, 
and the precision with which they can be used to deduce stellar
masses and radii, is surprising. 
The errors are notably larger when we replace either
the $f$- or the $w$-mode with the $p$-mode. For example, 
for
the combination (\ref{rfw}) and  (\ref{rpw}), the oscillation frequencies
of the $f$- and $p$-modes, the method estimates the stellar
parameters to within (23\%,123\%).
That is, from this combination we can at best get upper and lower bounds of
the parameters of the observed object.

\begin{table*}
 \centering
 \begin{minipage}{140mm}
  \caption{Results from a simple ``experiment'': We have used the
empirical relations for mode-frequencies of realistic EOS to deduce
the stellar parameters of a set of independent polytropic models.   
In the first three columns we list the parameters of the polytropes:
the polytropic index, the radius and the mass. The following columns
give
the percentage error in the estimated parameters (radius,mass) when
the polytropic mode-frequencies are used in our relations for 
$f$- and $w$-modes [(\ref{rfw}),(\ref{rftau}), 
(\ref{rww}) and (\ref{rwtau})]. We have not used the $p$-mode data
here, since the corresponding 
empirical relations are less robust than for the other modes.
}
  \begin{tabular}{@{}lrlcccccc@{}}
  $N$ & $R (km)$ & $M(M_\odot)$ & (\ref{rfw}) and  (\ref{rww})   
& (\ref{rfw}) and (\ref{rftau}) 
& (\ref{rfw}) and (\ref{rwtau}) 
& (\ref{rww}) and (\ref{rftau})  
& (\ref{rww}) and (\ref{rwtau}) & (\ref{rwtau}) and (\ref{rftau})\\
\hline
0.8 &10.03 & 1.08 &  (-3.5 , 0.1) &( -5.3 ,- 5.4) & (-3.9 ,-1.3)&(-2.6 ,-1.0) &(-2.2 ,-1.3) &( -2.8 , -1.3)\\
0.8 & 9.49 & 1.35 &  (-0.02,-1.8) &(  4.7 , 11.8) & ( 0.8 , 0.6)&(-4.6 ,-0.6) &(-2.9 ,-1.1) &( -1.5 , -0.5)\\
0.8 & 8.99 & 1.50 &  ( 1.3 ,-1.1) &(  0.1 , -5.1) & ( 2.1 , 1.1)&(-6.8 ,-1.6) &(-1.6 ,-1.4) &(  0.6 ,  0.02)\\ 
1.0 & 9.65 & 1.13 &  ( 3.4 ,-4.0) &(  9.7 , 15.2) & ( 4.4 ,-0.6)&(-0.5 ,-1.5) &(-0.3 ,-1.6) &( -0.5 , -1.7)\\
1.0 & 8.86 & 1.27 &  ( 6.6 ,-3.0) &( -3.5 ,-40.0) & ( 7.9 , 1.5)&(-1.7 ,-2.5) &( 1.9 ,-2.0) &( -0.8 ,-28.9)\\
1.0 & 7.42 & 1.35 &  ( 9.7 , 2.7) &( 10.9 ,  6.6) & (10.2 , 4.2)&(-4.4 ,-1.4) &( 7.7 , 1.9) &(  4.3 , -8.7)\\
1.2 &12.77 & 1.24 &  (-1.6 , 1.6) &(-12.0 ,-32.0) & (-2.5 ,-0.9)&( 3.0 ,-4.2) &( 0.7 ,-1.5) &(  4.4 , -1.8)\\
1.2 &10.48 & 1.44 &  ( 7.4 ,-3.5) &( -2.2 ,-38.7) & ( 7.8 ,-1.9)&( 3.8 ,-3.0) &( 5.6 ,-3.2) &( -3.2 , -4.7)\\
1.2 & 8.97 & 1.46 &  (11.4 , 0.3) &( 10.3 , -3.4) & (11.2 ,-0.2)&( 1.8 ,-1.5) &(12.0 , 0.5) &(  1.3 , -8.9)\\
\end{tabular}
\end{minipage}
\end{table*}
 
That the $p$-mode relation is less useful for an inversion to yield
the stellar mass and radius is, however, not completely bad news. 
Once we  have estimated the mass and the radius we want to
identify which of the proposed EOS that best fits the observed data.
When combined with data deduced from the other modes the $p$-modes
can provide the answer to this question, eg. via the results in
Figure~\ref{fptau}. If we  observe a $p$-mode we should at least
be able to exclude the unsuitable EOS.

\begin{figure}
\centerline{\epsfxsize=9cm \epsfysize=10cm \epsfbox{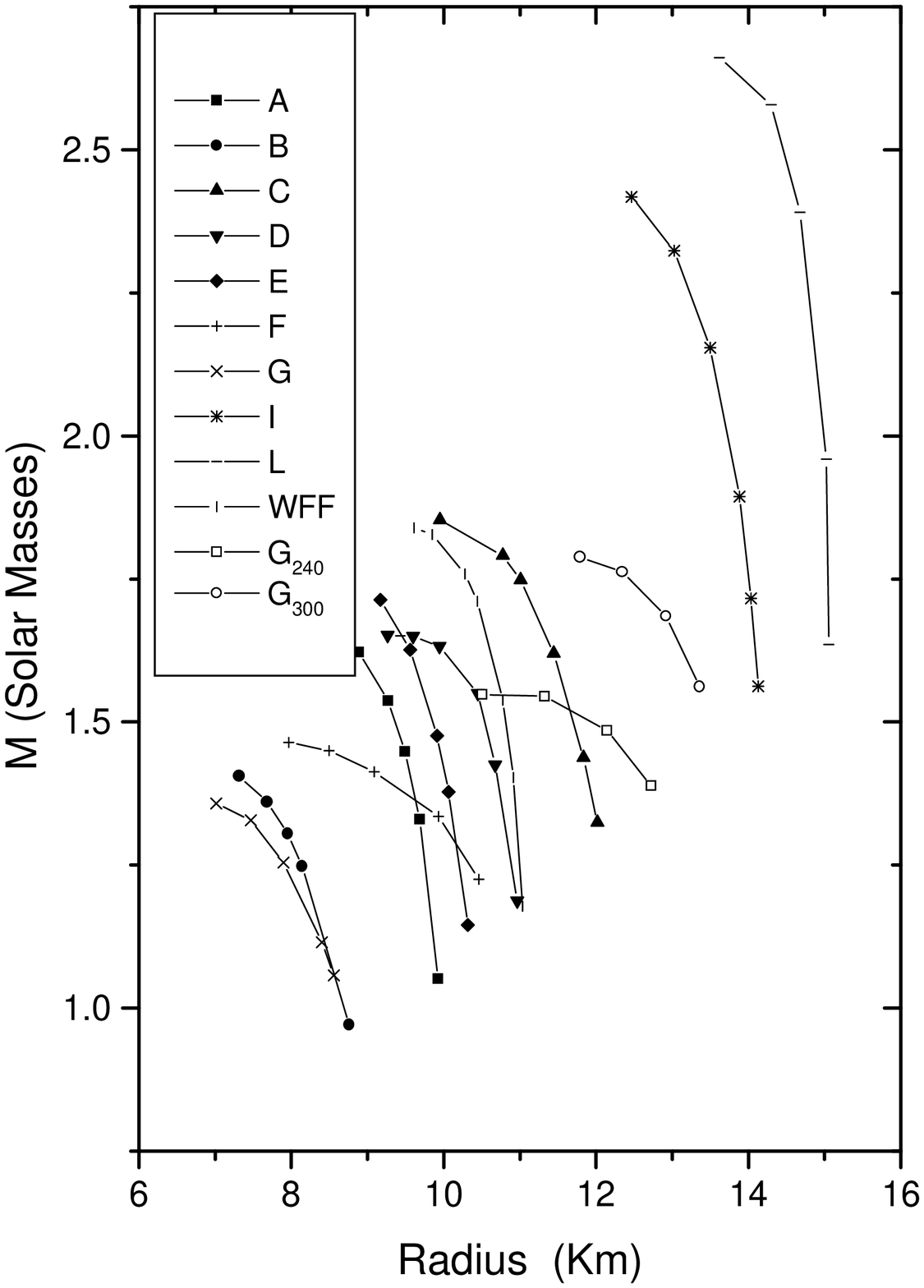}}
\caption{
The mass-radius relation for the twelve EOS in our sample. This data
can be used to identify which of  
the EOS that agrees best with the estimated parameters of a star.
From the graph one can easily identify the relative stiffness of the
EOS. In order of increasing stiffness they can be ordered  
$G < B < F < A < E < D < WFF < C < G_{240} < G_{300} < I <L$.
}
\label{fig6}
\end{figure}

The most suitable EOS can, of course, also be deduced from the
mass and radius of the star.    
As we show in Figure~\ref{fig6} the mass-radius relation 
is characteristic for each EOS in our sample.
From this data it should not be difficult to infer which EOS
that can lead to a mass and radius obtained via the empirical
relations.
Alternatively, one can use the approach suggested by Lindblom 
\shortcite{lind92}. He has shown how one can reconstruct the
 density-pressure relation in the interior of
a neutron star from a sample of observed masses and radii.

\section{Concluding remarks}

This paper concerns the feasibility of gravitational-wave
asteroseismology.
That is, whether it is realistic to expect to be able to 
infer stellar parameters [mass, radius and EOS]  
from observations of gravitational-wave signals from pulsating 
neutron stars.    
To address this issue we have calculated the details of
the modes that we expect to provide the strongest
gravitational waves: the $f$-mode,  the first $p$-mode and the
first $w$-mode, for a sample of twelve realistic EOS. 
Our chosen EOS span a range of stiffness that should include all 
proposed models. From the obtained data we have 
deduced a set of empirical relations that can be used to 
infer the mass and the radius (or rather, combinations thereof)
from observed mode-frequencies. In the ideal case, when both
the $f$- and the $w$-mode are detected, our empirical relations
predict $M$ and $R$ with surprising precision.

These results are undoubtedly interesting, but so far we have discussed
an ideal scenario. In order for gravitational-wave asteroseismology
to become reality, the detectors that are presently under
construction or at the planning stage must be able to detect
these mode-signals. We have argued that this may be possible in a
hand-waving
way, but we have as yet no detailed results. The answer depends to a
large
extent on how much energy is radiated through the pulsation modes of,
for example, a nascent neutron star immediately following the 
collapse. At the present time there are no reliable predictions of
this, and fully relativistic collapse simulations are desperately
needed. This is a challenge for numerical relativity, and hopefully
the answer will be known in the near future.    
Another, related issue, that we have not addressed
concerns  the expected detection errors. When the signal is noisy,
as is likely to be the case, there will be statistical errors
associated with the observed mode-data. These errors will obviously
affect the precision with which the stellar parameters can be deduced
from our empirical relations.  

Finally, the detectors which will be in operation in the 
next decade will mainly be  sensitive to frequencies 
below 5-6 kHz. The $f$-modes are well inside this bandwidth, 
as is the first $p$-mode for most of the EOS we have considered.
But only for the stiffest EOS does the first $w$-mode have a 
frequency around 6 kHz. This is bad news for the proposed
parameter detection/inversion. But as 
we have seen, a detection of the $w$-mode
could lead to a very accurate deduction of the stellar
mass and radius. The astrophysical pay-off would thus be
considerable, and the situation illustrates the need to complement
the currently planned detectors with ones dedicated to a
search for high-frequency gravitational waves.  

\section*{Acknowledgments}

It is a pleasure to thank N.K. Glendenning, P. Haensel and N. Stergioulas
for providing us with  data and information about the EOS.  
We are grateful to G. Allen, 
T. Apostolatos and B.F. Schutz for helpful discussions.
We also thank P. Laguna and P. Papadopoulos
for helping us optimize our numerical codes.  
This work was support by NATO research grant CRG960260.


%
%

\appendix

\section{Results for various Equations of State}

This appendix  provides the numerical data for mode-frequencies of 12
realistic EOS. This data was used to infer the empirical relations
discussed in the main body of the paper.
We provide the data in the form of one table for each EOS.
In each table we list the central density, the radius and the mass of the 
stellar model and the frequencies and the damping times
of the $f$-mode, the first $p$-mode and the first $w$-mode.

\begin{table*}
 \centering
 \begin{minipage}{140mm}
  \caption{Data for the EOS A, (Pandharipande 1971)}
  \begin{tabular}{@{}lllllllrl@{}}
 $\rho_c \times 10^{15}$       & $R$        & $M$  
 & $\omega_f$   & $\tau_f$   & $\omega_{p}$   & $\tau_{p}$
 & $\omega_{w_0}$ & $\tau_{w}$ \\
   $gr/cm^3$       & $km$        & $M_\odot$  
 & kHz   & s   & kHz   & s
 & kHz & ms  \\
3.980 & 8.426 & 1.653 & 3.090 & 0.109 & 7.838 & 4.640 &  9.824 & 0.064 \\ 
3.000 & 8.884 & 1.620 & 2.888 & 0.106 & 7.822 & 2.475 & 10.165 & 0.045 \\ 
2.344 & 9.268 & 1.535 & 2.704 & 0.109 & 7.819 & 2.163 & 10.766 & 0.032 \\  
1.995 & 9.493 & 1.447 & 2.579 & 0.117 & 7.818 & 2.293 & 11.444 & 0.027\\ 
1.698 & 9.688 & 1.328 & 2.447 & 0.132 & 7.807 & 2.726 & 12.344 & 0.022 \\
1.259 & 9.924 & 1.050 & 2.203 & 0.183 & 7.543 & 5.218 & 14.328 & 0.017\\   
\end{tabular}
\end{minipage}
\end{table*}


\begin{table*}
 \centering
 \begin{minipage}{140mm}
  \caption{Data for the EOS B, (Pandharipande 1971)}
  \begin{tabular}{@{}lllllllrl@{}}
 $\rho_c \times 10^{15}$       & $R$        & $M$  
 & $\omega_f$   & $\tau_f$   & $\omega_{p}$   & $\tau_{p}$
 & $\omega_{w}$ & $\tau_{w}$ \\
   $gr/cm^3$       & $km$        & $M_\odot$  
 & kHz   & s   & kHz   & s
 & kHz & ms \\
 5.012 &   7.317 & 1.405 & 3.598 & 0.091 & 8.957 & 2.994 & 11.577 & 0.053 \\
 7.684 &   7.682 & 1.360 & 3.393 & 0.089 & 8.739 & 1.615 & 12.094 & 0.037 \\
 3.388 &   7.951 & 1.303 & 3.236 & 0.091 & 8.515 & 1.406 & 12.638 & 0.029\\
 3.000 &   8.143 & 1.248 & 3.113 & 0.095 & 8.314 & 1.403 & 13.183 & 0.025 \\
 1.995 &   8.761 & 0.971 & 2.661 & 0.144 & 7.467 & 1.635 & 15.770 & 0.015 \\
\end{tabular}
\end{minipage}
\end{table*}

\begin{table*}
 \centering
 \begin{minipage}{140mm}
  \caption{Data for the EOS C, (Bethe \& Johnson 1974) (model I)}
  \begin{tabular}{@{}lllllllrl@{}}
 $\rho_c \times 10^{15}$       & $R$        & $M$  
 & $\omega_f$   & $\tau_f$   & $\omega_{p}$   & $\tau_{p}$
 & $\omega_{w}$ & $\tau_{w}$ \\
   $gr/cm^3$       & $km$        & $M_\odot$  
 & kHz   & s   & kHz   & s
 & kHz & ms  \\
3.0  &  9.952 & 1.852 & 2.656 & 0.121 & 6.432 & 1.548 &  8.843 & 0.062 \\ 
1.995& 10.778 & 1.790 & 2.383 & 0.124 & 6.209 & 1.452 &  9.214 & 0.040\\ 
1.778& 11.009 & 1.746 & 2.304 & 0.129 & 6.126 & 1.514 &  9.444 & 0.035\\ 
1.413& 11.441 & 1.619 & 2.144 & 0.146 & 5.925 & 1.766 & 10.124 & 0.028 \\ 
1.122& 11.832 & 1.435 & 1.975 & 0.181 & 5.664 & 2.264 & 11.003 & 0.023\\  
1.0  & 12.017 & 1.322 & 1.885 & 0.208 & 5.505 & 2.741 & 11.496 & 0.021\\  
\end{tabular}
\end{minipage}
\end{table*}


\begin{table*}
 \centering
 \begin{minipage}{140mm}
  \caption{Data for the EOS D,   (Bethe \& Johnson 1974) (model V)}
  \begin{tabular}{@{}lllllllrl@{}}
 $\rho_c \times 10^{15}$       & $R$        & $M$  
 & $\omega_f$   & $\tau_f$   & $\omega_{p}$   & $\tau_{p}$
 & $\omega_{w}$ & $\tau_{w}$  \\
   $gr/cm^3$       & $km$        & $M_\odot$  
 & kHz   & s   & kHz   & s
 & kHz & ms \\
3.548 &  9.262 & 1.652 & 2.833 & 0.108 & 6.646 & 0.822 &  9.954 & 0.049   \\
3.0   &  9.597 & 1.649 & 2.699 & 0.110 & 6.662 & 0.900 & 10.001 & 0.043   \\ 
2.512 &  9.945 & 1.632 & 2.560 & 0.114 & 6.456 & 1.106 & 10.114 & 0.037  \\
1.778 & 10.448 & 1.549 & 2.356 & 0.127 & 5.911 & 1.763 & 10.543 & 0.029  \\
1.413 & 10.678 & 1.425 & 2.235 & 0.147 & 5.881 & 1.945 & 11.371 & 0.024   \\
1.122 & 10.968 & 1.187 & 2.045 & 0.194 & 5.351 & 2.455 & 12.792 & 0.019   \\ 
\end{tabular}
\end{minipage}
\end{table*}


\begin{table*}
 \centering
 \begin{minipage}{140mm}
  \caption{Data for the EOS E, (Moszkowski 1974)}
  \begin{tabular}{@{}lllllllrl@{}}
 $\rho_c \times 10^{15}$       & $R$        & $M$  
 & $\omega_f$   & $\tau_f$   & $\omega_{p}$   & $\tau_{p}$
 & $\omega_{w}$ & $\tau_{w}$  \\
   $gr/cm^3$       & $km$        & $M_\odot$  
 & kHz   & s   & kHz   & s
 & kHz & ms  \\
2.818 &  9.171 & 1.713 & 2.805 & 0.109 & 7.553 & 3.503 &  9.593 & 0.053 \\
2.239 &  9.562 & 1.626 & 2.642 & 0.111 & 7.474 & 2.372 & 10.175 & 0.038 \\
1.778 &  9.915 & 1.476 & 2.467 & 0.123 & 7.327 & 2.256 & 11.170 & 0.027\\
1.585 & 10.066 & 1.378 & 2.376 & 0.135 & 7.212 & 2.476 & 11.812 & 0.024\\
1.259 & 10.316 & 1.145 & 2.180 & 0.179 & 6.865 & 3.404 & 13.321 & 0.018 \\
\end{tabular}
\end{minipage}
\end{table*}


\begin{table*}
 \centering
 \begin{minipage}{140mm}
  \caption{Data for the EOS F, (Arponen 1972)}
  \begin{tabular}{@{}lllllllrl@{}}
 $\rho_c \times 10^{15}$       & $R$        & $M$  
 & $\omega_f$   & $\tau_f$   & $\omega_{p}$   & $\tau_{p}$
 & $\omega_{w}$ & $\tau_{w}$  \\
   $gr/cm^3$       & $km$        & $M_\odot$  
 & kHz   & s   & kHz   & s
 & kHz & ms  \\
5.012 &  7.966 & 1.464 & 3.403 & 0.097 & 7.349 & 0.826 & 11.123 & 0.0551\\
3.981 &  8.495 & 1.450 & 3.138 & 0.098 & 7.087 & 0.631 & 11.318 & 0.0409\\
3.162 &  9.087 & 1.413 & 2.859 & 0.104 & 6.855 & 0.601 & 11.560 & 0.0316 \\
2.239 &  9.934 & 1.335 & 2.478 & 0.130 & 6.585 & 0.948 & 12.011 & 0.0235 \\
1.585 & 10.462 & 1.225 & 2.231 & 0.157 & 6.361 & 1.653 & 12.655 & 0.0194\\
1.122 & 10.892 & 1.031 & 1.980 &  0.475 & 6.039 & 3.194 & 13.705 & 0.0167 \\
\end{tabular}
\end{minipage}
\end{table*}


\begin{table*}
 \centering
 \begin{minipage}{140mm}
  \caption{Data for the EOS G, (Canuto \& Chitre 1974) }
  \begin{tabular}{@{}lllllllrl@{}}
 $\rho_c \times 10^{15}$       & $R$        & $M$  
 & $\omega_f$   & $\tau_f$   & $\omega_{p}$   & $\tau_{p}$
 & $\omega_{w}$ & $\tau_{w}$ \\
   $gr/cm^3$       & $km$        & $M_\odot$  
 & kHz   & s   & kHz   & s
 & kHz & ms  \\
6.042 & 7.010 & 1.356 & 3.801 & 0.091 & 9.029 & 1.994 & 11.931 & 0.056  \\  
4.503 & 7.472 & 1.327 & 3.526 & 0.087 & 8.722 & 1.223 & 12.402 & 0.037\\ 
3.498 & 7.898 & 1.253 & 3.264 & 0.091 & 8.387 & 1.024 & 13.146 & 0.026 \\ 
2.631 & 8.397 & 1.114 & 2.927 & 0.111 & 7.897 & 1.194 & 14.556 & 0.019\\  
2.376 & 8.556 & 1.057 & 2.813 & 0.123 & 7.705 & 1.360 & 15.069 & 0.017\\

\end{tabular}
\end{minipage}
\end{table*}

\begin{table*}
 \centering
 \begin{minipage}{140mm}
  \caption{Data for the EOS I, (Cohen et al. 1970)}
  \begin{tabular}{@{}lllllllrl@{}}
 $\rho_c \times 10^{15}$       & $R$        & $M$  
 & $\omega_f$   & $\tau_f$   & $\omega_{p}$   & $\tau_{p}$
 & $\omega_{w}$ & $\tau_{w}$  \\
   $gr/cm^3$       & $km$        & $M_\odot$  
 & kHz   & s   & kHz   & s
 & kHz & ms  \\
1.585  & 12.468 & 2.418 & 2.063 & 0.158 & 5.444 & 7.729 & 6.758 & 0.083\\  
1.259  & 13.023 & 2.324 & 1.943 & 0.154 & 5.415 & 3.796 & 7.100 & 0.058 \\   
1.0000 & 13.498 & 2.154 & 1.822 & 0.163 & 5.358 & 3.270 & 7.685 & 0.042 \\
0.7943 & 13.885 & 1.894 & 1.689 & 0.193 & 5.255 & 3.736 & 8.571 & 0.032 \\
0.6310 & 14.128 & 1.562 & 1.551 & 0.255 & 5.087 & 5.294 & 9.686 & 0.025  \\ 

\end{tabular}
\end{minipage}
\end{table*}

\begin{table*}
 \centering
 \begin{minipage}{140mm}
  \caption{Data for the EOS L, (Pandharipande et al. 1976)}
  \begin{tabular}{@{}lllllllrl@{}}
 $\rho_c \times 10^{15}$       & $R$        & $M$  
 & $\omega_f$   & $\tau_f$   & $\omega_{p}$   & $\tau_{p}$
 & $\omega_{w}$ & $\tau_{w}$ \\
   $gr/cm^3$       & $km$        & $M_\odot$  
 & kHz   & s   & kHz   & s
 & kHz & ms \\
1.500  & 13.617 & 2.661 &  1.874 & 0.173 & 5.099 & 7.716 &  6.142 & 0.092\\ 
1.259  & 13.935 & 2.649 &  1.816 & 0.168 & 5.123 & 7.829 &  6.204 & 0.079 \\ 
1.000  & 14.297 & 2.579 &  1.746 & 0.170 & 5.147 & 8.204 &  6.401 & 0.063\\ 
0.794  & 14.678 & 2.391 &  1.655 & 0.183 & 5.179 & 5.451 &  6.940 & 0.047 \\  
0.631  & 14.986 & 2.044 &  1.534 & 0.216 & 5.215 & 6.067 &  8.007 & 0.034\\  
0.600  & 15.022 & 1.959 &  1.508 & 0.229 & 5.215 & 6.552 &  8.259 & 0.032\\   
0.500  & 15.053 & 1.636 &  1.415 & 0.312 & 5.152 & 9.755 &  9.224 & 0.026\\   
0.398  & 14.885 & 1.214 &  1.303 &11.494 & 4.865 &17.094 & 10.629 & 0.020\\ 
\end{tabular}
\end{minipage}
\end{table*}


\begin{table*}
 \centering
 \begin{minipage}{140mm}
  \caption{Data for the EOS WFF, (Wiringa  et al. 1988)}
  \begin{tabular}{@{}lllllllrl@{}}
 $\rho_c \times 10^{15}$       & $R$        & $M$  
 & $\omega_f$   & $\tau_f$   & $\omega_{p}$   & $\tau_{p}$
 & $\omega_{w}$ & $\tau_{w}$  \\
   $gr/cm^3$       & $km$        & $M_\odot$  
 & kHz   & s   & kHz   & s
 & kHz & ms \\
4.0   &  9.177 & 1.827 & 2.854 & 0.123 & 7.010 & 3.949 & 8.824 & 0.081\\
3.0   &  9.612 & 1.840 & 2.695 & 0.120 & 7.000 & 3.014 & 8.893 & 0.062\\
2.6   &  9.849 & 1.828 & 2.609 & 0.119 & 7.003 & 2.743 & 8.993 & 0.054 \\
2.0   & 10.277 & 1.759 & 2.449 & 0.121 & 7.016 & 2.478 & 9.383 & 0.040 \\
1.8   & 10.440 & 1.710 & 2.382 & 0.124 & 7.016 & 2.504 & 9.662 & 0.035\\
1.4   & 10.774 & 1.538 & 2.216 & 0.142 & 6.976 & 3.050 &10.726 & 0.026\\
1.216 & 10.912 & 1.403 & 2.118 & 0.163 & 6.911 & 3.628 &11.506 & 0.023\\
1.0   & 11.036 & 1.178 & 1.977 & 0.203 & 6.745 & 5.004 &12.767 & 0.019\\
0.9   & 11.073 & 1.044 & 1.897 & 1.927 & 6.609 & 6.214 &13.547 & 0.017 \\
0.8   & 11.101 & 0.889 & 1.804 & 5.598 & 6.387 & 9.275 &14.514 & 0.015\\
\end{tabular}
\end{minipage}
\end{table*}

\begin{table*}
 \centering
 \begin{minipage}{140mm}
  \caption{Data for the EOS G$_{240}$, (Glendenning 1985) }
  \begin{tabular}{@{}lllllllrl@{}}
 $\rho_c \times 10^{15}$       & $R$        & $M$  
 & $\omega_f$   & $\tau_f$   & $\omega_{p}$   & $\tau_{p}$
 & $\omega_{w}$ & $\tau_{w}$ \\
   $gr/cm^3$       & $km$        & $M_\odot$  
 & kHz   & s   & kHz   & s
 & kHz & ms\\
2.515 & 10.907 & 1.553 & 2.346 & 0.134 & 5.456 & 0.505 & 10.417 & 0.030 \\ 
1.889 & 11.531 & 1.536 & 2.140 & 0.153 & 5.289 & 0.704 & 10.406 & 0.027\\
1.429 & 12.146 & 1.485 & 1.942 & 0.183 & 5.163 & 1.086 & 10.515 & 0.025\\
1.088 & 12.651 & 1.405 & 1.774 & 0.221 & 5.079 & 1.809 & 10.719 & 0.023\\
0.762 & 13.148 & 1.240 & 1.581 & 4.965 & 4.971 & 3.975 & 11.205 & 0.020 \\
0.587 & 13.338 & 1.079 & 1.464 &10.812 & 4.842 & 6.998 & 11.752 & 0.018 \\
\end{tabular}
\end{minipage}
\end{table*}

\begin{table*}
 \centering
 \begin{minipage}{140mm}
  \caption{Data for the EOS G$_{300}$, (Glendenning 1985)}
  \begin{tabular}{@{}lllllllrl@{}}
 $\rho_c \times 10^{15}$       & $R$        & $M$  
 & $\omega_f$   & $\tau_f$   & $\omega_{p}$   & $\tau_{p}$
 & $\omega_{w}$ & $\tau_{w}$\\
   $gr/cm^3$       & $km$        & $M_\odot$  
 & kHz   & s   & kHz   & s
 & kHz & ms  \\
2.063 & 11.790 & 1.788 & 2.151 & 0.141 & 5.278 & 0.713 &  9.145 & 0.036  \\
1.543 & 12.343 & 1.762 & 1.991 & 0.157 & 5.196 & 0.984 &  9.225 & 0.032 \\
1.162 & 12.920 & 1.685 & 1.818 & 0.186 & 5.157 & 1.556 &  9.489 & 0.028\\
0.883 & 13.362 & 1.562 & 1.669 & 0.227 & 5.152 & 2.724 &  9.878 & 0.025\\
0.645 & 13.681 & 1.345 & 1.512 & 0.812 & 5.127 & 5.832 & 10.581 & 0.022\\
0.518 & 13.736 & 1.156 & 1.419 & 4.894 & 5.029 & 9.772 & 11.256 & 0.019\\
\end{tabular}
\end{minipage}
\end{table*}

\end{document}